# Identification of fake stereo audio

Tianyun Liu[0000-0003-0056-3447] and Diqun Yan[0000-0002-5241-7276]

College of Information Science and Engineering, Ningbo University, Zhejiang Ningbo
315211, China
`yandiqun@nbu.edu.cn`

**Abstract.** Channel is one of the important criteria for digital audio quality. Generally, stereo audio two channels can provide better perceptual quality than mono audio. To seek illegal commercial benefit, one might convert mono audio to stereo one with fake quality. Identifying of stereo faking audio is still a less-investigated audio forensic issue. In this paper, a stereo faking corpus is first present, which is created by Haas Effect technique. Then the effect of stereo faking on Mel Frequency Cepstral Coefficients (MFCC) is analyzed to find the difference between the real and faked stereo audio. Finally, an effective algorithm for identifying stereo faking audio is proposed, in which 80-dimensional MFCC features and Support Vector Machine (SVM) classifier are adopted. The experimental results on three datasets with five different cut-off frequencies show that the proposed algorithm can effectively detect stereo faking audio and achieve a good robustness.

## 1 Introduction

Audio forensics [1] is an important branch of multimedia security, which can be used to evaluate the authenticity of digital audio. Many audio forensic methods have been proposed for various speech operations [2, 3]. In addition to common audio forgeries such as pitch-shifting [4, 5, 6], device source [7, 8, 9], replaying [10] and the detection of the operation type and sequence of digital speech [11]. Fake-quality detection is a very important part of the field of audio forensics, such as the detection of fake quality [12]. The so-called fake quality refers to recompress low bit rate audio into high bit rate audio. Generally, higher bit rate indicates better audio quality. Since bit rate increasing will cause recompression, the author in [12] proposed a fake quality detection algorithm based on double compression trace. As far as we know, however, there is no work related to stereo faking detection, which also belongs to the category of fake-quality.

Stereo faking is one of audio quality faking technique, which aims to convert a mono audio into a stereo one. Mono audio is a single channel of sound perceived as coming from one position. Stereo audio is the reproduction of sound using two or more independent audio channels in a way that creates the impression of sound heard from various directions, as in natural hearing. Stereo audio has almost completely replaced mono because of the improved audio quality. Since the 1960s to 1970s and even earlier, due to the limitations of recording equipment, most of audio recordings and movie soundtracks were monophonic, and their auditory effects were poor. To improve the audio quality, one can create an artificial stereo audio from its mono version. With the development of the audio edit software/tools, the quality of created stereo audio can be very close to real stereo audio. This technique, however, is also easily grasped by some criminals for illegal benefits. For example, people can buy their favorite songs from Internet. Often these songs are in stereo format. The legitimate interests of consumers will be violated if the bought songs are fake stereo created from mono audio.

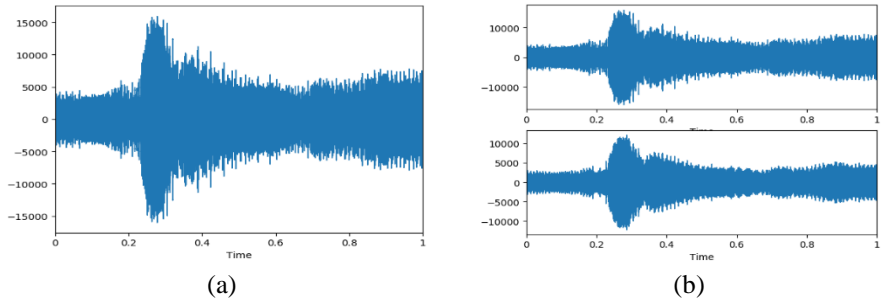

**Fig. 1.** Audio waveforms. (a) Real mono audio (b) Fake stereo audio

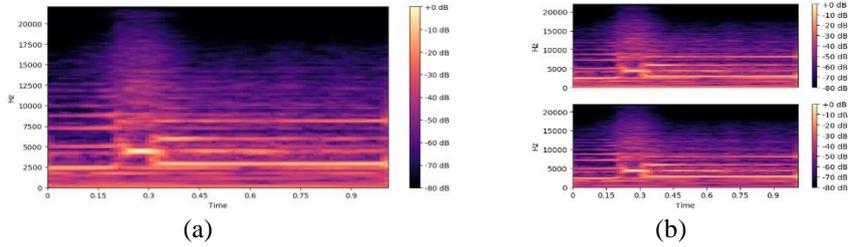

**Fig. 2.** Audio spectrograms. (a) Real mono audio (b) Fake stereo audio

As shown in Fig. 1 and Fig. 2, it is a hard task to distinguish fake stereo audio from the real one with the waveform or spectrogram, when the mono audio which is the source of the fake stereo audio is not given. However, when we listen to these fake stereo audios, we can feel a significant decline in sound quality. Hence, it is necessary to design a detection algorithm for fake stereo audio. In this paper, a corpus for fake stereo audio detection is created from three real stereo datasets. Then we propose an effective algorithm for detecting fake stereo audio for detecting fake stereo audio, which is based on MFCC [13] features and Support Vector Machine (SVM) classifier. Considering on the information of audio sources and channels, six various classification models are trained. The experimental results show that the proposed algorithm can achieve a good performance on detecting the fake stereo audio. The main contributions of this paper are as follows:

• To our knowledge, this is the first forensic work to identify fake stereo audio in the field of fake quality forensics, provides a new kind of content for the field of audio forensics.

• We provide a corpus for fake stereo audio detection. The corpus contains a public stereo audio dataset. At the same time, we have collected a large number of samples from the two most widely used fields (music and film recording) as datasets.

• After studying the method of stereo forgery, we found that the cut-off frequency of different high-pass filters has little effect on the sound quality but can greatly affect the detection effect of the model. The detection algorithm we put forward based on this point has strong robustness, and can detect fake stereo audios with different cut-off frequencies of high-pass filters.

• Through our research on the distribution of three different stereo audio samples in their feature space, we provide a general detection model. The model can effectively detect fake stereo audio in three different scenarios, that is, the model trained under the music dataset.

The rest of this paper is organized as follows. In Section 2, the detail of the fake stereo corpus is given. Section 3 describes the proposed detection algorithm. In Section 4, we present the experimental results in various cases. Finally, the conclusions are given in Section 5.

## 2  Stereo faking corpus

For the purposes of stereo faking identification, a corpus of various stereo audio is required. A new fake stereo audio corpus was established which consists of three stereo forms as described below.

### 2.1  Stereo Faking

The aim of stereo faking is to convert mono audio into an artificial stereo. There are two typical methods to create fake stereo audio. One method called Channel Copy. First, a faked channel can be made by coping the original channel of the mono audio straight. Then by integrating these two channels, the fake stereo audio will be obtained. The creation of fake stereo audio with Channel Copy can be expressed by (1), (2) and (3).

$$X = x_l, l = 1,2,3, \dots, L \tag{1}$$

$$\hat{x}_l = x_l, l = 1,2,3, \dots, L \tag{2}$$

$$Y = \{x_l, \hat{x}_l\} = y_{n,l} = \begin{cases} y_{n,l} = x_l, & n = 1 \\ y_{n,l} = \hat{x}_l, & n = 2 \end{cases} \tag{3}$$

where $X$ stands for mono audio, $x$ is the audio channel data, $L$ is the length of the audio file, $Y$ represents fake stereo audio file, $n$ denotes the number of channels. It is easy to identify the fake stereo created by Channel Copy. The suspected stereo audio can be determined as a fake stereo audio if its two channels are the same ($\hat{x}_l = x_l$).

In realistic scenarios, the attacker would take more advanced methods to create the fake stereo audio. Haas Effect [14] is the other method to create more deceptive fake stereo audio, which can be expressed by (4), (5) and (6),

$$X = x_l, l = 1,2,3,\dots,L \qquad (4)$$

$$\hat{x}_l = F(x_l) \neq x_l, l = 1,2,3,\dots,L \qquad (5)$$

$$Y = \{x_l, \hat{x}_l\} = y_{n,l} = \begin{cases} y_{n,l} = x_l, & n = 1 \\ y_{n,l} = \hat{x}_l, & n = 2 \end{cases} \qquad (6)$$

where $F$ denotes the high-pass filter to high-pass the forged channel object. The specific implementation of the high-pass filter $F$ in this work can be defined as (7),

$$\hat{x}_l = \frac{1}{1+2\pi f_{cut-off}T}(x_l - x_{l-1} + \hat{x}_{l-1}) \qquad (7)$$

where $T$ symbolizes the cycle and $f_{cof}$ is the cut-off frequency (cof).

## 2.2   Real and faking stereo dataset

The real stereo audio in the corpus comes from the following three sources: DCASE2019 [15] recordings, IMDbTop250 [16] movies and QQ Music [17] songs.

DCASE is the challenge on Detection and Classification of Acoustic Scenes and Events providing a great opportunity for development and comparison of state-of-the-art methods. The DCASE2019 dataset consists of real stereo audio which includes 8 scenes such as airports, stations, parks, streets, etc., and each clip is a stereo audio with a duration of 10 seconds. In this work, 275 clips are randomly selected from each scene. Finally, 2200 clips from DCASE2019 are taken to constitute the first real stereo dataset.

IMDb (Internet Movie Database) is an online database about movie actors, movies, TV shows, TV stars and movie productions. Ten movies of IMDb Top250, which are Shawshank's Redemption, Hachiko, the Jurassic Park, Seven Deadly Sins, Dream Stealing Space, Truman, Forrest Gump, Return of Tarzan, Detective Chinatown 2, Number one player, are considered in this work. The soundtracks of these movies are extracted first, and then each of them are segmented into 30 minutes. Therefore, the 2nd real stereo dataset named FILM consists a total 5 hours audio clips.

QQ Music is a leading music streaming platform in China. It provides tens of millions of songs. A total of 91 songs in QQ Music are downloaded as the 3rd real stereo dataset named MUSIC.

Each clip in real stereo datasets are segmented into 1 second and used to create fake stereo clip by Haas Effect. Finally, we constructed a corpus for fake stereo detection. The corpus is divided into training and testing sets as shown in Table 1.

**Table 1.** Information of three major datasets

| Corpus | Training datasets | Testing datasets |
| --- | --- | --- |
| DCASE2019 | 15998 | 5999 |
| MUSIC | 16684 | 5895 |
| FILM | 12600 | 5400 |

According to the description in subsection A, the cut-off frequency of the high-pass filter is a very critical parameter in the Haas Effect. In order to evaluate the performance of the algorithm comprehensively, two kinds of audio channels and five kinds of cut-off frequencies in high-pass filter are taken into consideration. Specially, the cut-off frequency for the training set is fixed at 200 Hz, and the cut-off frequencies for the testing set are configured to 200 Hz, 400 Hz, 600 Hz, 800 Hz, and 1000 Hz, respectively. The depth bits of the three datasets are all 16 bits, and the formats are all wav formats. The sampling rate of the DCASE dataset is 48000 Hz, and the sampling rate of the remaining two datasets is 44100 Hz.

## 3  Identification of fake stereo audios

In this section, we first introduce the extraction of MFCC features. Then the SVM classifier used as classification is briefly described. Finally, the identification algorithm of fake stereo audio is present.

### 3.1  Feature extraction

In MFCC is the most widely used acoustic feature for speech recognition in the frequency domain [13]. It models the spectral energy distribution in a perceptually meaningful way. Before calculating the MFCC, in order to amplify the high-frequency components, a finite impulse response (FIR) high-pass filter should be used to pre-emphasize the input speech $X = [x(1), x(2), \ldots, x(i), \ldots x(I)]$ as follows,

$$x'(i) = x(i) - 0.95x(i-1) \qquad (8)$$

where $I$ is the size of the input audio.

Then, by multiplying each frame by the Hamming window, the emphasis signal $x'(i)$ is divided into overlapped frames. The Hamming window is,

$$w(n) = 0.54 - 0.46 \cos\left(\frac{2\pi n}{N-1}\right), 0 \leq n \leq N-1 \quad (9)$$

where $N$ is the length of the window.

Next, the frequency spectrum is obtained by applying Fast Fourier Transform (FFT) on each windowed frame. Then, the spectrum is decomposed into multiple sub-bands using a set of triangular Mel-scale bandpass filters. Let $E(b), 0 \leq b \leq B$ represents the sum of the power spectral coefficients in the $bth$ sub-bands, where $B$ is the total number of filters,

The MFCC can be calculated by applying the discrete cosine transform (DCT) to the logarithm of $E(b)$ as follows,

$$C(l) = \sum_{b=0}^{B-1} \log_{10}(1 + E(b)) \cos(l\frac{\pi}{B}(b + 0.5)), 0 \leq l \leq L \quad (10)$$

where $L$ is the length of the MFCC. In this experiment, we extracted the MFCC features of the left and right channels of the stereo to be tested separately, and stitched the features of the right channel to the left channel, and finally we get $2L$ dimensional features.

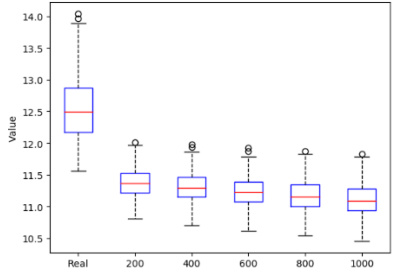

(a)

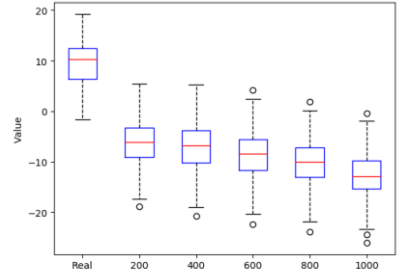

(b)

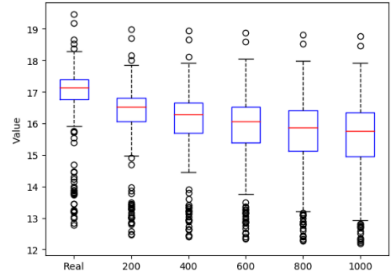

(c)

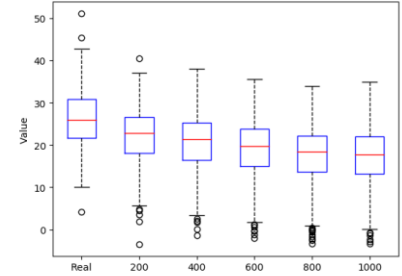

(d)

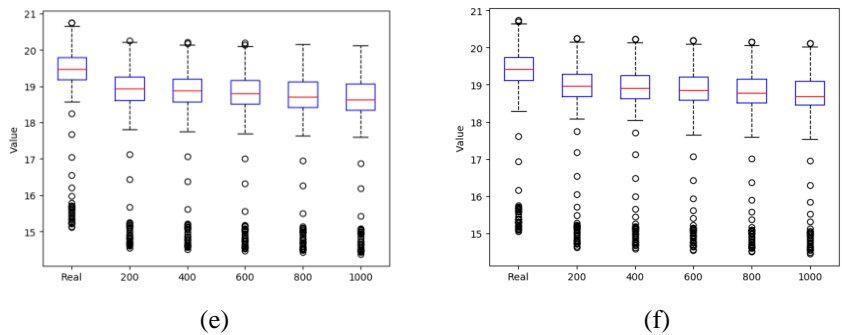

|  |  |
|:---:|:---:|
| (e) | (f) |

**Fig. 3.** Distributions of the MFCC components from the real and fake stereo audios (F200 denotes the $f_{cut-off}$ is 200). (a) 1st component of right channel in DCASE2019 (b) 2nd component of left channel in DCASE2019 (c) 1st component of right channel in FILM (d) 2nd component of left channel in FILM (e) 1st component of right channel in MUSIC (f) 2nd component of left channel in MUSIC

In Fig. 3, the boxplots of the MFCC features from the real and fake stereo audios are shown. Various cut-off frequencies of high-pass filtering are considered. It can be seen that the distributions of the MFCC features in DCASE2019 have an obvious change between the real and fake stereo audios. In FILM and MUSIC datasets, such change still exists although it is not obvious. Since the MFCC features can expose the effect caused by stereo faking, they are used to identify the fake stereo.

### 3.2 Algorithm

The proposed identification algorithm is based on MFCC statistical moments and SVM classifiers. SVM is a supervised learning technique that is a powerful discriminative classifier and it can be used in fake stereo audio detection. The classification decision function of SVM can be defined as,

$$f(x) = sign(\sum_{i=1}^{N} \alpha_i y_i K(x_i, x_j) + b) \quad (11)$$

where $y_i$ are the real output values, $\sum_{i=1}^{N} \alpha_i y_i = 0$ and $\alpha_i > 0$. The support vectors $x_j$, their corresponding weights $\alpha_i$ and bias $b$ are calculated on the training set. The kernel function $K(x_i, x_j)$ can be defined as $K(x, y) = \langle x, y \rangle$, this is actually the inner product of the original input space. The kernel function is to replace the inner product of the feature space with the inner product of the input vector, so as to achieve the purpose of mapping the input space data to a high-dimensional space. In high-dimensional space, the two classes are easier to separate with hyperplane. Finally, the sigmoid ($sign$) activation function is used to get the final classification result.

As shown in Fig. 4. Real and fake stereo based on left channel in Fig. 4 means that the real mono audio is used as the left channel of fake stereo audio, and the right channel is faked. Real and fake stereo based on right channel means the same. Considering that forensics cannot know which channel the forger will use for forgery, it is necessary to train the left and right channels separately. Therefore, we have two SVM models for each dataset, called 1st SVM classifier and 2nd SVM classifier.

Our algorithm is divided into the training and testing stages. In the training stage, the dataset for training is composed of a real stereo set and two fake stereo sets (1st SVM classifier and 2nd SVM classifier). MFCC coefficients are extracted from the training set as the detection features. The features are used as the training features to train two SVM classifiers. And each SVM classifier can be used to identify whether a testing stereo audio is faked or not. In the testing stage, the MFCC features from real and fake stereo audio in test set are extracted. Then, they used as the input into the trained SVM classifiers. These two classification results are combined to obtain a final detection result. If all the two results are real, the testing stereo audio is identified as a real one. If at least one result is fake, the testing stereo audio is identified as a fake one.

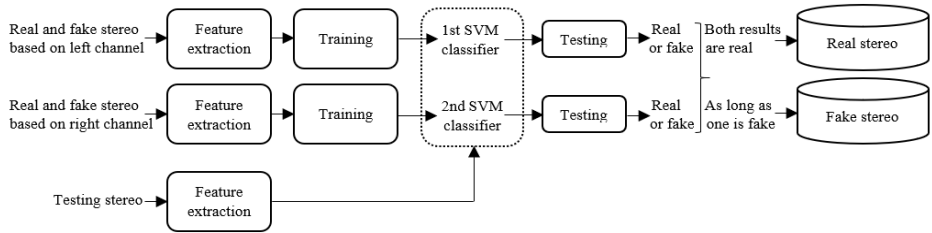

**Fig. 4.** Proposed identification algorithm of fake stereo.

## 4    Experimental results

In this section, we will present the experimental results of the proposed method. First, the experimental setups for feature extraction, classifier and corpus are described. Then the experimental results for intra-dataset and cross-dataset are given to evaluate the performance of the proposed method.

### 4.1    Experimental setup

In our experiment, fake stereo audio is created by Audacity with Haas Effect. Each dataset in the corpus is divided into two disjoint parts by the ratio of 6: 4. The first 60% is the training set, and the remaining 40% is used for verification. For the extraction of its MFCC features, $L$ is 40 and $2L$ is 80. For the SVM classifier, the penalty coefficient $C$ set to 0.4, and the linear kernel function is adopted.

### 4.2 Intra-dataset evaluation

This case means that the dataset for testing is the same as that for training. In this link, our detection accuracy rate (ACC) can reach up to 99%. Table 2 shows the identification performance of the proposed method. It can be seen from Table 2 that the false acceptance rate (FAR) is almost close to 0. It indicates that the method's performance is remarkable.

**Table 2.** FAR for intra-dataset evaluation

|   | DCASE | FILM | MUSIC |
|---|---|---|---|
| L | 0.02 | 0.02 | 0.08 |
| R | 0.05 | 0.06 | 0.12 |

### 4.3 Various high-pass filter parameters for testing

In this experiment, two classification models are trained with left and right channels at the cut-off frequency 200 Hz. The stereo audio faked with various cut-off frequencies are tested. The experimental results for left and right channels are shown in Table 3 and Table 4, respectively. It can be seen that the performance can be kept at a high and stable level with the increase of cut-off frequency.

**Table 3.** ACC and FAR of the left channel model with various $f_{cof}$ (%)

| Cut-off frequency (Hz) | DCASE2019 | FILM | MUSIC |
|---|---|---|---|
| 200 | 99/0.02 | 99/0.02 | 99/0.08 |
| 400 | 99/0.00 | 100/0.00 | 99/0.00 |
| 600 | 99/0.00 | 100/0.00 | 99/0.02 |
| 800 | 99/0.00 | 100/0.00 | 99/0.02 |
| 1000 | 99/0.00 | 100/0.00 | 99/0.00 |

**Table 4.** ACC and FAR of the left channel model with various $f_{cof}$ (%)

| Cut-off frequency (Hz) | DCASE2019 | FILM | MUSIC |
|---|---|---|---|
| 200 | 99/0.02 | 99/0.02 | 99/0.08 |
| 400 | 99/0.00 | 99/0.00 | 99/0.00 |
| 600 | 99/0.00 | 99/0.00 | 99/0.00 |
| 800 | 99/0.00 | 99/0.00 | 99/0.00 |
| 1000 | 99/0.00 | 99/0.06 | 99/0.00 |

In this experiment, two classification models are trained with left and right channels at the cut-off frequency 1000 Hz. The stereo audio faked by various cut-off frequencies are tested. The experimental results for left and right channels are shown in Fig. 5. It can be seen that, with the decrease of cut-off frequency from 1000 to 200 Hz, the performance of the model tends to decline, especially at 600Hz, its performance declines especially.

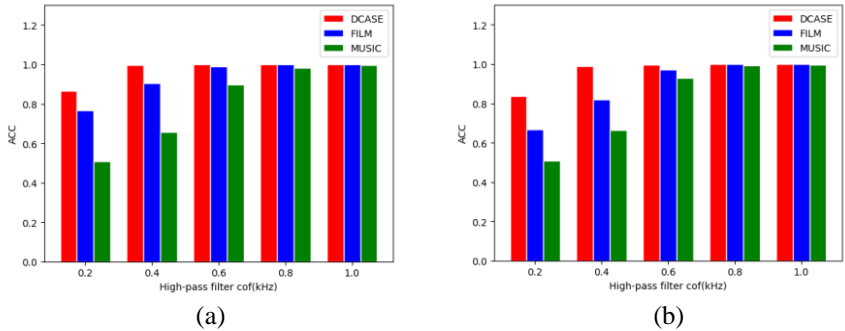

**Fig. 5.** Experimental results of the left and right channel model with various $f_{cof}$ (ACC). (a) 1st SVM classifier trained by 1000Hz high-pass filter $f_{cof}$. (b) 2nd SVM classifier trained by 1000Hz high-pass filter $f_{cof}$.

### 4.4 Cross-dataset evaluation

In the practical forensic scenarios, the channels of suspected audio are variable in recording devices and environments. In our corpus, DCASE2019 is environmental recording, FILM is movie recording, and MUSIC comes as songs. Their recording equipment and environment are also different from each other. Hence, cross-database evaluation is a necessary and important issue. Similar to subsections B and C, when using a dataset for training, the stereo through all three datasets are tested. Note that the datasets used for training and testing are different.

The detection performance of the cross-dataset evaluation is shown in Table 5 and 6. Compared with Table 2, the cross-dataset performance is a little worse than the intra-dataset case. In particular, the model trained under the FILM dataset has an accuracy rate of less than 60% when detecting the MUSIC test set, and the FAR is higher than 0.8. The accuracy of the model trained under the DCASE2019 dataset detects the FILM and MUSIC test set is less than 90%. However, it can be seen that most detection rates are above 95%, while FAR is below 7%, especially, the model trained under the MUSIC dataset shows good performance and strong robustness in all three test sets. Representative that our proposed algorithm is still effective to identify fake stereo audio.

Table 5. ACC and FAR for the 1st SVM classifier (%).

|  | DCASE2019 | FILM | MUSIC |
|---|---|---|---|
| DCASE2019 | 99/0.02 | 89/0.04 | 86/14.00 |
| FILM | 97/5.00 | 99/0.02 | 55/88.00 |
| MUSIC | 99/0.00 | 99/0.00 | 99/0.08 |

Table 6. ACC and FAR for the 2nd SVM classifier (%).

|  | DCASE2019 | FILM | MUSIC |
|---|---|---|---|
| DCASE2019 | 99/0.05 | 85/0.26 | 85/21.00 |
| FILM | 96/6.00 | 99/0.06 | 54/86.00 |
| MUSIC | 99/0.00 | 99/0.00 | 99/0.12 |

## 5   Conclusion

In this paper, an algorithm for identifying stereo faking audio is proposed. Statistical analysis of the MFCC features shows that the distribution of the feature components of real stereo audio changes due to the forgery of the channel. An identification algorithm based on MFCC features and SVM classifiers is proposed in this paper. In the experiments, three stereo audio datasets are used for intra-dataset and cross-dataset evaluations. The experimental results show a good performance in the case of various high-pass filter cut-off frequencies. It indicates that the proposed algorithm can effectively identify stereo faking audio.